\documentclass[aps,graphicx,twocolumn,prd,eqsecnum,showpacs]{revtex4-1}
\usepackage{graphicx}
\usepackage{dcolumn}
\usepackage{bm}
\begin{document}

\title{Electromagnetic form factor of pion in the field theory inspired approach.}
\author{N.~N.~Achasov}
\email[]{achasov@math.nsc.ru} \affiliation{Laboratory of
Theoretical Physics, S.~L.~Sobolev Institute for Mathematics,
630090, Novosibirsk, Russian Federation
}%
\author{A.~A.~Kozhevnikov}
\email[]{kozhev@math.nsc.ru} \affiliation{Laboratory of
Theoretical Physics, S.~L.~Sobolev Institute for Mathematics, and
Novosibirsk State University, 630090, Novosibirsk, Russian
Federation}

\date{\today}
\begin{abstract}
A new expression for the pion form factor $F_\pi$ is proposed. It
takes into account the pseudoscalar meson loops and the mixing of
$\rho(770)$ with heavier $\rho(1450)$ and $\rho(1700)$ resonances.
The expression has correct analytical properties and can be used
in both timelike and spacelike kinematical regions. The comparison
is made  with the existing experimental data on $F_\pi$ collected
with the detectors SND, CMD-2, KLOE, and the BaBaR restricted to
energies below 1 GeV. A good description of all four data sets is
obtained. In the  spacelike region,  upon substituting the
resonance parameters found in the timelike one, one obtains
$F_\pi$ in agreement with the measurements of NA7 Collaboration.

\end{abstract}

\maketitle

\section{introduction}
\label{intro} ~

The pion form factor $F_\pi$ is an important characteristics of
the low energy phenomena in particle physics  related with the
hadronic properties of the electromagnetic current in the
theoretical scheme of the vector dominance model
\cite{Sak,gell61,KLZ,Lowen}. There are a number of expressions for
this quantity used in the analysis of experimental data. The
simplest approximate vector dominance model  expression based on
the effective $\gamma-\rho$ coupling $\propto\rho_\mu A_\mu$
\cite{KLZ},
\begin{equation}
F_\pi(s)=\frac{m^2_\rho g_{\rho\pi\pi}/g_\rho}
{m^2_\rho-s-i\sqrt{s}\Gamma_{\rho\pi\pi}(s)},\label{Fpisingle}\end{equation}
(for notations see Sec.~\ref{formula}) does not possess the
correct analytical properties upon the continuation to the
unphysical region $0\leq s<4m^2_\pi$ and further to the spacelike
region $s\leq0$, nor does it takes into account the mixing of the
isovector $\rho$-like resonances. Since, phenomenologically,
\cite{pdg} $g_{\rho\pi\pi}/g_\rho$ is not equal to unity$-$to be
precise,
\begin{equation}
\frac{g_{\rho\pi\pi}}{g_\rho}=\left(\frac{3m_\rho\Gamma_{\rho\pi\pi}
\Gamma_{\rho ee}}{2\alpha^2q^3_{\pi}}\right)^{1/2}\approx1.20
\label{normFpi}
\end{equation}
$-$the correct normalization $F_\pi(0)=1$ is satisfied by
Eq.~(\ref{Fpisingle}) only approximately. Hereafter,
$\alpha=1/137$ stands for the fine structure constant. The formula
of Gounaris and Sakurai \cite{GS} respects the above normalization
condition and has the correct properties under analytical
continuation. However,  being based on some sort of  effective
radius approximation for the single $\rho(770)$ resonance, it is
not suited for taking into account the mixing of $\rho(770)$ with
heavier isovector mesons. The expression analogous to
Eq.~(\ref{Fpisingle}) based on the gauge invariant $\gamma-\rho$
coupling $\propto\rho_{\mu\nu}F_{\mu\nu}$,
\begin{equation}
F_\pi(s)=1+\frac{sg_{\rho\pi\pi}/g_\rho}
{m^2_\rho-s-i\sqrt{s}\Gamma_{\rho\pi\pi}(s)},\label{Fpigau}\end{equation}
respect the correct normalization, but does not possesses correct
analytical properties  and breaks unitarity. The  earlier
expression \cite{ach97,ach97a} for $F_\pi$  takes into account the
strong isovector mixing, but has the shortcoming that the above
normalization condition is satisfied only approximately, within
the accuracy 20$\%$.

The applications of the Lagrangian of Kroll, Lee, and Zumino
\cite{KLZ} to the calculations of $F_\pi$ with the meson loop
contributions in the field-theoretic context  are given, in
particular, in Refs.~\cite{Kapusta,Dom07,Dom08}. In particular,
Ref.~\cite{Dom07} contains the comparison of the theoretical
$F_\pi$ with the experimental data in the spacelike kinematical
region. However, the authors of Ref.~\cite{Dom07} refrained from
the application of their expression  in the timelike region
despite the fact that the high statistics experimental data
collected with the detectors SND \cite{snd} and CMD-2 \cite{cmd}
were available at that time.

The purpose of the present work is to obtain the expression  for
the pion form factor which possesses the correct analytical
properties  in the entire kinematic domain and takes into account
the mixing of $\rho(770)$ with the heavier resonances $\rho(1450)$
and $\rho(1700)$. By restricting the consideration to the
inclusion of the pseudoscalar meson loops $\pi^+\pi^-$ and $K\bar
K$, which admits the analytical treatment and is valid at energies
below 1 GeV, the new expression is found and compared with the
existing data on $F_\pi$ collected with the detectors SND
\cite{snd} CMD-2 \cite{cmd}, KLOE \cite{kloe}, and BaBaR
\cite{babar}.

Below, in Sec.~\ref{prelim}, the method is described by which the
loop contributions to the vector$-$meson propagators are taken
into account.  The expression for the form factor $F_\pi(s)$  is
given in Sec.~\ref{formula}. Section~\ref{analysis} is devoted to
the analysis of available new experimental data on $F_\pi(s)$
\cite{snd,cmd,kloe,babar} . Section~\ref{discussion} contains the
discussion of the obtained results. The conclusions are stated in
Sec.~\ref{conclusion}. The Appendix is devoted to the description
of the method by which  the resonance mixing is taken into
account.

\section{The loop contributions to the vector meson propagator}
\label{prelim}~

Let us give some details necessary for the derivation of the
expression for the pion form factor. They refer to the
pseudoscalar loop contributions.   For the sake of brevity, the
notation
\begin{equation} \rho_1\equiv\rho(770)\mbox{,
}\rho_2\equiv\rho(1450)\mbox{,
}\rho_3\equiv\rho(1700)\label{rho123}\end{equation} is used
hereafter  for the isovector resonances involved in the
consideration.

\begin{figure}
\includegraphics[width=60mm]{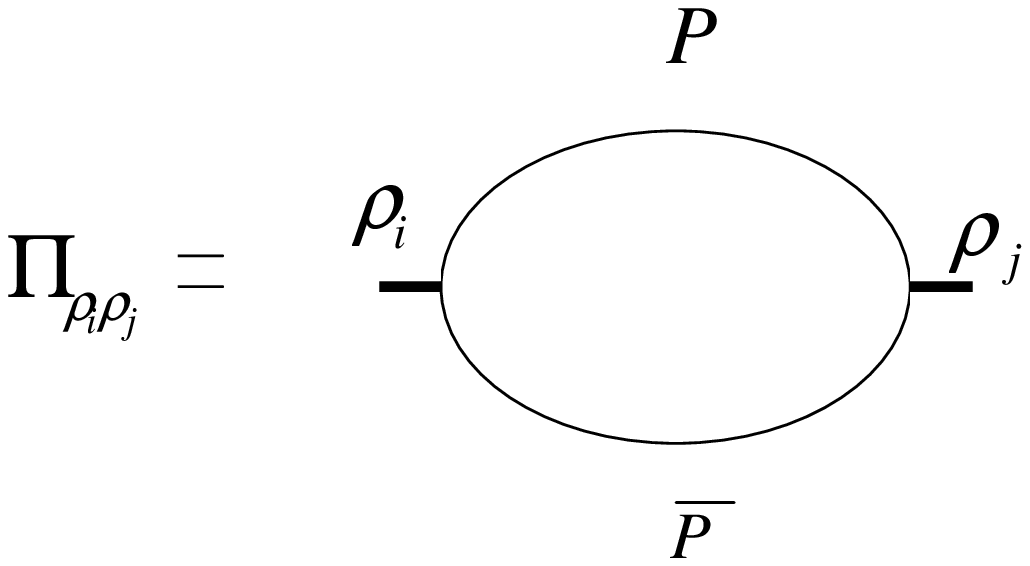}
\caption{\label{polPP}The meson loop diagram contributing  to both
the diagonal polarization  operator $\Pi_{\rho_i\rho_i}-$
resulting, in particular, in the finite width of the resonance$-$
and the nondiagonal one $\Pi_{\rho_i\rho_j}$, responsible for the
$\rho_i\rho_j$ resonance mixing;  $P=\pi^+,K^+,K^0$.}
\end{figure}

The starting point is the effective Lagrangian describing the
SU(3) invariant  interaction of the vector resonances with the
pair of pseudoscalar  mesons \cite{gellsu3,Neem}.  Restricted to
the couplings of the isovector resonances $\rho_i$, $i=1,2,3$,
with the pair of pions and kaons ($P=\pi,K$), this Lagrangian
looks like
\begin{eqnarray}
{\cal
L}_{\rho_iPP}&=&ig_{\rho_i\pi\pi}\rho^0_{i\mu}\left\{\frac{1}{2}\left[K^-\partial_\mu
K^+-K^+\partial_\mu
K^--\right.\right.\nonumber\\&&\left.\left.\bar K^0\partial_\mu
K^0+K^0\partial\bar
K^0\right]+\pi^-\partial_\mu\pi^+\right.\nonumber\\&&\left.-\pi^+\partial_\mu\pi^-
\right\}.\label{LRpipi}\end{eqnarray}  The partial width of the
decay $\rho_i\to P\bar P$ calculated from the above effective
Lagrangian, is
\begin{equation}
\Gamma_{\rho_i\to
PP}(s)=\frac{g^2_{\rho_iPP}s^{1/2}v_P^3(s)}{48\pi
},\label{Gam2pi}\end{equation} where $s$ stands for the (virtual)
mass squared of the decaying resonance $\rho_i$, and
\begin{equation}
v_P(s)=\sqrt{1-\frac{4m^2_P}{s}}\label{vps}\end{equation} is the
velocity of the final meson in the rest frame of the decaying
resonance. Applying the Cutkosky cutting rule to the diagram in
Fig.~\ref{polPP} one finds that the
imaginary part of the diagonal polarization operator caused by the
specific real intermediate  state $P\bar P$ is related to the
corresponding partial decay width according to the expression
\begin{equation}
{\rm Im}\Pi^{P\bar
P}_{\rho_i\rho_i}(s)=\sqrt{s}\Gamma_{\rho_iPP}(s).\label{ImPiii}\end{equation}
In the present work, the real intermediate states $\pi^+\pi^-$,
$K^+K^-$, and $K^0\bar K^0$ are taken into account, hence
$${\rm Im}\Pi_{\rho_i\rho_i}(s)=\sum_{P=\pi^+,K^+,K^0}{\rm Im}\Pi^{P\bar P}_{\rho_i\rho_i}(s).$$

The diagonal and nondiagonal  polarization operators for the
specific loop $P\bar P$ are calculated from the dispersion
integral. Here the version of this integral is defined which
automatically provides the condition $\Pi_{\rho_i\rho_j}(0)=0$, in
agreement with the conservation of the vector current. To this
end, the dispersion relation should be written for the quantity
$\Pi_{\rho_i\rho_j}(s)/s$. Then, one has
\begin{eqnarray}
\frac{\Pi^{P\bar
P}_{\rho_i\rho_j}(s)}{s}&=&\frac{1}{\pi}\int_{4m^2_P}^\infty\frac{{\rm
Im}\Pi^{P\bar
P}_{\rho_i\rho_j}(s^\prime)ds^\prime}{s^\prime(s^\prime-s-i\varepsilon)}=
\nonumber\\&&\frac{g_{\rho_iPP}g_{\rho_iPP}}{48\pi^2}
\int_{4m^2_P}^\infty\frac{v^3_P(s^\prime)ds^\prime}{s^\prime-s-i\varepsilon}
.\label{disprel}\end{eqnarray} One can evaluate this dispersion
integral in the unphysical region $0\leq s<4m^2_P$, where ${\rm
Im}\Pi_{\rho_i\rho_j}=0$, and  no pole is encountered. But,  the
integral is still divergent  at $s^\prime\to\infty$. The
divergence can be regularized by taking the cutoff $s^\prime_{\rm
max}=\Lambda^2$. The integration can be fulfilled with the change
of the integration variable
$\sigma^2=v^2_P(s^\prime)=1-4m^2_P/s^\prime$:
\begin{eqnarray*}
I(s)&\equiv&\int_{4m^2_P}^{\Lambda^2}\frac{ds^\prime}{s^\prime-s}
\left(1-\frac{4m^2_P}{s^\prime}\right)^{3/2}=
\int_0^{1-2m^2_P/\Lambda^2}d\sigma\times\nonumber\\&&
\frac{8m^2_P\sigma^4}{(1-\sigma^2)(4m^2_P-s+\sigma^2)}
=-\frac{8m^2_P}{s}+\nonumber\\&&
2\left(\frac{4m^2_P}{s}-1\right)^{3/2}\arctan\frac{1}{\sqrt{\frac{4m^2_P}{s}-1}}
+4\ln\frac{\Lambda}{m_P}.
\end{eqnarray*}
The logarithmic divergence can be removed by fixing
Re$I(m^2_V)=0$.  The diagonal elements
$\Pi_{\rho_i\rho_i}\equiv\Pi_{\rho_i\rho_i}(s)$ can be represented
in the form
\begin{eqnarray}
\Pi_{\rho_i\rho_i}&=&\frac{sg^2_{\rho_i\pi\pi}}{48\pi^2}
\left[\Pi_\pi(s,m^2_{\rho_i})+
\frac{1}{2}\Pi_K(s,m^2_{\rho_i})\right],\label{Piii}\end{eqnarray}
where the factor 1/2 in the second term is due to the flavor SU(3)
relation $g_{\rho_iKK}=\frac{1}{2}g_{\rho_i\pi\pi}$ [see
Eq.~(\ref{LRpipi})] and that two isotopic $K\bar K$ modes
contribute.

The expressions for $\Pi_{\pi,K}(s,m^2_V)$ are represented in the
following form. Since the pion is the lightest hadron, the
function $\Pi_\pi(s,m^2_V)$ looks as
\begin{widetext}
\begin{eqnarray}
\Pi_\pi(s,m^2_V)&=&8m^2_\pi\left(\frac{1}{m^2_V}-\frac{1}{s}\right)+v^3_\pi(m^2_V)
\ln\frac{1+v_\pi(m^2_V)}{1-v_\pi(m^2_V)}+v^3_\pi(s)\left[i\pi-\ln\frac{1+v_\pi(s)}{1-v_\pi(s)}\right]
\mbox{, if }s\geq4m^2_\pi;\nonumber\\
\Pi_\pi(s,m^2_V)&=&8m^2_\pi\left(\frac{1}{m^2_V}-\frac{1}{s}\right)+v^3_\pi(m^2_V)
\ln\frac{1+v_\pi(m^2_V)}{1-v_\pi(m^2_V)}+2\bar
v^3_\pi(s)\arctan\frac{1}{\bar v_\pi(s)}\mbox{, if }0\leq
s<4m^2_\pi;\nonumber\\
\Pi_\pi(s,m^2_V)&=&8m^2_\pi\left(\frac{1}{m^2_V}-\frac{1}{s}\right)+v^3_\pi(m^2_V)
\ln\frac{1+v_\pi(m^2_V)}{1-v_\pi(m^2_V)}-v^3_\pi(s)\ln\frac{v_\pi(s)+1}{v_\pi(s)-1}
\mbox{, if }s<0. \label{Pipi}
\end{eqnarray}
\end{widetext}
The function $\Pi_K(s,m^2_V)$ looks different depending on the
mass of the vector meson $m_V$. If $m_V>2m_K$, as is the case for
$V=\rho(1450)$ and $\rho(1700)$, the expression is
\begin{widetext}
\begin{eqnarray}
\Pi_K(s,m^2_V)&=&8m^2_K\left(\frac{1}{m^2_V}-\frac{1}{s}\right)+v^3_K(m^2_V)\ln\frac{1+v_K(m^2_V)}{1-v_K(m^2_V)}
+v^3_K(s)\left[i\pi-\ln\frac{1+v_K(s)}{1-v_K(s)}\right]\mbox{, if
}s\geq4m^2_K;\nonumber\\
\Pi_K(s,m^2_V)&=&8m^2_K\left(\frac{1}{m^2_V}-\frac{1}{s}\right)+v^3_K(m^2_V)\ln\frac{1+v_K(m^2_V)}{1-v_K(m^2_V)}
+2\bar v^3_K(s)\arctan\frac{1}{\bar v_K(s)}\mbox{, if } 0\leq
s<4m^2_K;\nonumber\\
\Pi_K(s,m^2_V)&=&8m^2_K\left(\frac{1}{m^2_V}-\frac{1}{s}\right)+v^3_K(m^2_V)\ln\frac{1+v_K(m^2_V)}{1-v_K(m^2_V)}
-v^3_K(s)\ln\frac{v_K(s)+1}{v_K(s)-1}\mbox{, if }
s<0.\label{PiKK23}\end{eqnarray}
\end{widetext}
If $m_V<2m_K$, as is the case for $V=\rho(770)$, the expression is
\begin{widetext}
\begin{eqnarray}
\Pi_K(s,m^2_V)&=&8m^2_K\left(\frac{1}{m^2_V}-\frac{1}{s}\right)-2\bar
v^3_K(m^2_V)\arctan\frac{1}{\bar v_K(m^2_V)}
+v^3_K(s)\left[i\pi-\ln\frac{1+v_K(s)}{1-v_K(s)}\right]\mbox{, if
}s\geq4m^2_K;\nonumber\\
\Pi_K(s,m^2_V)&=&8m^2_K\left(\frac{1}{m^2_V}-\frac{1}{s}\right)-2\bar
v^3_K(m^2_V)\arctan\frac{1}{\bar v_K(m^2_V)}+2\bar
v^3_K(s)\arctan\frac{1}{\bar v_K(s)}\mbox{, if } 0\leq
s<4m^2_K;\nonumber\\
\Pi_K(s,m^2_V)&=&8m^2_K\left(\frac{1}{m^2_V}-\frac{1}{s}\right)-2\bar
v^3_K(m^2_V)\arctan\frac{1}{\bar
v_K(m^2_V)}-v^3_K(s)\ln\frac{v_K(s)+1}{v_K(s)-1}\mbox{, if }
s<0.\label{PiKK1}\end{eqnarray}
\end{widetext}
The function $v_P(s)$ ($P=\pi,K$) is given by Eq.~(\ref{vps}),
while
\begin{equation}
\bar v_P(s)=\sqrt{\frac{4m^2_P}{s}-1}.\label{vbarps}\end{equation}
Note that the expressions Eqs.~(\ref{Pipi}), (\ref{PiKK23}), and
(\ref{PiKK1}) have the property that their real parts vanish at
$s=m^2_V$:
$${\rm Re}\Pi_{\pi,K}(m^2_V,m^2_V)=0.$$

\section{The expression  for the pion form factor}
\label{formula} ~

The new expression for the  pion form factor which automatically
respects the current conservation condition $F_\pi(0)=1$ and
possesses the correct analytical properties over the entire  $s$
axis, looks like
 \begin{eqnarray}
F_\pi(s)&=&(g_{\gamma\rho_1},g_{\gamma\rho_2},g_{\gamma\rho_3})G^{-1}\left(%
\begin{array}{c}
  g_{\rho_1\pi\pi} \\
  g_{\rho_2\pi\pi} \\
  g_{\rho_3\pi\pi} \\
\end{array}%
\right)+\nonumber\\&&
\frac{g_{\gamma\omega}\Pi_{\rho_1\omega}}{D_\omega\Delta}\left(g_{11}g_{\rho_1\pi\pi}+
g_{12}g_{\rho_2\pi\pi}+\right.\nonumber\\&&
\left.g_{13}g_{\rho_3\pi\pi}\right). \label{Fpifin}\end{eqnarray}
The notations are as follows. The quantity
\begin{equation}
g_{\gamma V}=\frac{m^2_V}{g_V},\label{garhg}\end{equation}
($V=\rho_{1,2,3},\omega$) is introduced in such a way that
$eg_{\gamma V}$, where $e$ is the electric charge, is the $\gamma
V$ transition amplitude. As usual, the coupling constant $g_V$ is
calculated from the electronic width
\begin{equation}
\Gamma_{V\to
e^+e^-}=\frac{4\pi\alpha^2m_V}{3g^2_V}\label{gamee}\end{equation}
of the resonance $V$. The matrix of inverse propagators
\begin{equation}
G=\left(%
\begin{array}{ccc}
  D_{\rho_1} & -\Pi_{\rho_1\rho_2} & -\Pi_{\rho_1\rho_3}  \\
  -\Pi_{\rho_1\rho_2} & D_{\rho_2} & -\Pi_{\rho_2\rho_3}  \\
  -\Pi_{\rho_1\rho_3} & -\Pi_{\rho_2\rho_3} & D_{\rho_3}  \\
  \end{array}%
\right) \label{G}\end{equation} is responsible for the
$\rho(770)$-$\rho(1450)$-$\rho(1700)$ mixing
\cite{ach84,ach84a,ach97,ach97a,ach02}, and $\Delta={\rm det}G$.
See the Appendix for more detail. The inverse propagators of the
$\rho_i-$resonance ($i=1,2,3$) are
\begin{equation}
D_{\rho_i}=m^2_{\rho_i}-s-\Pi_{\rho_i\rho_i},\label{Drhoi}\end{equation}
where  the diagonal  polarization operator $\Pi_{\rho_i\rho_i}$
can be expressed through the  functions $\Pi_\pi(s,m^2_V)$ and
$\Pi_K(s,m^2_V)$ described in Sec.~\ref{prelim}. The nondiagonal
polarization operators are the following:
\begin{eqnarray}
\Pi_{\rho_1\rho_2}&=&\frac{g_{\rho_2\pi\pi}}{g_{\rho_1\pi\pi}}\Pi_{\rho_1\rho_1},\nonumber\\
\Pi_{\rho_1\rho_3}&=&\frac{g_{\rho_3\pi\pi}}{g_{\rho_1\pi\pi}}\Pi_{\rho_1\rho_1},\nonumber\\
\Pi_{\rho_2\rho_3}&=&\frac{g_{\rho_2\pi\pi}g_{\rho_3\pi\pi}}{g^2_{\rho_1\pi\pi}}\Pi_{\rho_1\rho_1}
+sa_{23}. \label{Piij}\end{eqnarray} The quantity $a_{23}$ is the
dimensionless  phenomenological free parameter. No such parameter
is introduced in $\Pi_{\rho_1\rho_2}$ and $\Pi_{\rho_1\rho_3}$
because it would result in a shift of the $\rho(770)$ resonance
peak position. See the Appendix and  Refs.~\cite{ach97,ach97a}.

The term $\propto\Pi_{\rho_1\omega}$ in Eq.~(\ref{Fpifin}) takes
into account the $\rho(770)-\omega(782)$ mixing. The basic
quantities in this contribution are the following. The inverse
propagator of the meson $\omega(782)$ is taken in the form
\begin{equation}
D_\omega=m^2_\omega-s-i\sqrt{s}\Gamma_\omega,\label{Dom}\end{equation}
where the energy-dependent width
$$\Gamma_\omega\equiv\Gamma_\omega(s)=\Gamma_{\omega3\pi}(s)+\Gamma_{\omega\pi\gamma}(s)
+\Gamma_{\omega\eta\gamma}(s)$$ includes the dominant decay mode
$\omega(782)\to\pi^+\pi^-\pi^0$ and the radiative ones. The tree
pion decay width is represented in the form
$$\Gamma_{\omega3\pi}(s)=\frac{g^2_{\omega\rho_1\pi}}{4\pi}W_{3\pi}(s),$$
where $W_{3\pi}(s)$ is the phase space volume of the final
$\pi^+\pi^-\pi^0$ state:
\begin{eqnarray}
W_{3\pi}(s)&=&\int_{2m_\pi}^{\sqrt{s}-m_\pi}dmm^2\Gamma_{\rho_1\pi\pi}(m^2)q^3_{\rho\pi}
\int_{-1}^1dx\times\nonumber\\&&(1-x^2)\left|\frac{1}{D_{\rho_1}(m^2)}+\frac{1}{D_{\rho_1}(m^2_+)}
+\right.\nonumber\\&&\left.\frac{1}{D_{\rho_1}(m^2_-)}\right|^2.\end{eqnarray}
Here,  $m$ is the invariant mass of the $\pi^+\pi^-$ pair while
$m_\pm$ refers to the $\pi^\pm\pi^0$ one:
\begin{eqnarray}
m^2_\pm&=&\frac{1}{2}(s+3m^2_\pi-m^2)\pm\nonumber\\&&
xq_{\rho\pi}\sqrt{s\left(1-\frac{4m^2_\pi}{m^2}\right)},
\label{kinem}
\end{eqnarray}
and $q_{\rho\pi}=q(\sqrt{s},m,m_\pi)$.  Here and in what follows,
\begin{eqnarray}
q(\sqrt{s},m_a,m_b)&=&\frac{1}{2\sqrt{s}}
\{[s-(m_a+m_b)^2]\times\nonumber\\&&[s-(m_a-m_b)^2]\}^{1/2}
\label{qab}\end{eqnarray} is the momentum of the particles $a$ or
$b$ with the  masses  $m_a$ or $m_b$, respectively, in the rest
reference frame of the decaying particle whose  invariant mass is
$\sqrt{s}$. The coupling constant $g_{\omega\rho\pi}$ is evaluated
from the $\omega\to\pi^+\pi^-\pi^0$ decay width. The
energy-dependent radiative width $\Gamma_{VP\gamma}(s)$, where
$V=\rho_1,\omega$, $P=\pi,\eta$, is related to the radiative width
on the mass shell
$\Gamma^{(0)}_{VP\gamma}\equiv\Gamma_{VP\gamma}(m^2_V)$ in accord
with the relation
\begin{equation}
\Gamma_{VP\gamma}(s)=\Gamma^{(0)}_{VP\gamma}\frac{q^3_P(s)}{q^3_P(m^2_V)},\label{Gamrad}\end{equation}
and $q_P(s)=q(\sqrt{s},m_P,0)$ is the momentum of the
pseudoscalar meson $P$ in the rest frame of the decaying vector
meson $V$. The quantity
\begin{eqnarray}
\Pi_{\rho_1\omega}&=&\frac{s}{m^2_\omega}\Pi_{\rho_1\omega}^\prime+
i\sqrt{s\Gamma_{\omega\pi\gamma}(s)\Gamma_{\rho\pi\gamma}(s)}
\label{Pirhom}\end{eqnarray} is the polarization operator of the
$\rho(770)-\omega(782)$ mixing. The real part
$s\Pi_{\rho_1\omega}^\prime/m^2_\omega$ is chosen in such a way
that it vanishes at $s=0$, and $\Pi_{\rho_1\omega}^\prime$ is a
free parameter. The  contributions to Im$\Pi_{\rho_1\omega}$ from
the $\eta\gamma$ intermediate state can be neglected in comparison
with the $\pi\gamma$ one.  If not fitted, the masses and partial
widths of particles and resonances involved in the treatment   are
taken from the Review of Particle Physics \cite{pdg}.

Note that  the isovector-isoscalar type of weak  mixing is
essential only for the $\rho(770)-\omega(782)$ system because it
is enhanced due to the small mass difference of these resonances.
As for other isovector-isoscalar mixings $\rho(1450)-\omega(782)$
and $\rho(1700)-\omega(782)$, there is no enhancement due to the
mass proximity, and    one can  neglect $\Pi_{\rho_{2,3}\omega}$
in what follows. The coupling constant of the direct transition
$\omega\to\pi^+\pi^-$ is neglected, too. The reason for this is
explained in the Appendix. See Eq.~(\ref{gom2pi}) and the
discussion around it. The quantities $g_{11}$, $g_{12}$, $g_{13}$
are, respectively,
\begin{eqnarray*}
g_{11}&=&D_{\rho_2}D_{\rho_3}-\Pi^2_{\rho_2\rho_3},\nonumber\\
g_{12}&=&D_{\rho_3}\Pi_{\rho_1\rho_2}+\Pi_{\rho_1\rho_3}\Pi_{\rho_2\rho_3},\nonumber\\
g_{13}&=&D_{\rho_2}\Pi_{\rho_1\rho_3}+\Pi_{\rho_1\rho_2}\Pi_{\rho_2\rho_3}.
\end{eqnarray*}
See Eq.~(\ref{gij}) in the Appendix.

When checking the form factor normalization $F_\pi(0)=1$, one
should have in mind that the $\rho\omega$ mixing  is negligible at
$s=0$, because, at this energy squared,   there is no enhancement
of the effect due to the proximity of $m_\omega$ and $m_\rho$. The
same is true for other contributions violating G-parity
conservation. Neglecting the above contributions  results in the
correct normalization $F_\pi(0)=1$,  if one takes
\begin{equation}
\frac{g_{\rho_1\pi\pi}}{g_{\rho_1}}+\frac{g_{\rho_2\pi\pi}}{g_{\rho_2}}+
\frac{g_{\rho_3\pi\pi}}{g_{\rho_3}}=1.\label{norm}\end{equation}
Indeed,  the  mixings due to strong interactions
$\Pi_{\rho_i\rho_j}$ vanish at $s=0$, and $F_\pi(0)$ reduces to
the above sum. This is the reason for the $s$ in front of $a_{23}$
in Eq.~(\ref{Piij}). The comparison of the new expression
Eq.~(\ref{Fpifin}) with the latest experimental data
\cite{snd,cmd,kloe,babar} obtained in $e^+e^-$ annihilation is
presented in the next section.

\section{The data analysis and results}
\label{analysis} ~

The experimental data on the reaction $e^+e^-\to\pi^+\pi^-$
collected by  the collaborations SND \cite{snd}, CMD-2 \cite{cmd},
KLOE \cite{kloe}, and BaBaR \cite{babar}   are chosen for the
analysis in the framework of the field-theory-inspired approach to
the pion form factor presented in this work.  As for the BaBaR
data set, we restrict ourselves by the points with $\sqrt{s}\leq1$
GeV, because, at the first stage of the study, the proposed
expression for the polarization operator is restricted to include
only $\pi^+\pi^-$ and $K\bar K$ loops.

The original $e^+e^-\to\pi^+\pi^-$ data of the SND, CMD-2, and
KLOE Collaborations are presented in two distinct  forms. The
first one is the form factor with the vacuum polarization effect
included. The BaBaR Collaboration does not present their results
in this form. The second form is the so-called bare cross section.
This quantity is undressed from the vacuum polarization effects,
but includes  the final state radiation. All four groups present
their data in this form. For the purpose of uniformity of
presentation, the analysis of the present work refers to the bare
cross section
\begin{equation}
\sigma_{\rm
bare}=\frac{8\pi\alpha^2}{3s^{5/2}}|F_\pi(s)|^2q^3_\pi(s)
\left[1+\frac{\alpha}{\pi}a(s)\right],\label{sigbare}
\end{equation}
where $F_\pi(s)$ is given by Eq.~(\ref{Fpifin}),
$$q_\pi(s)=\sqrt{s}v_\pi(s)/2$$ is the momentum of the final pion, and
the function $a(s)$ allows for the radiation of a photon by the
final pions. In the case of the pointlike pions, it has the form
\cite{snd,schwing,deers,melnik,hoef}
\begin{widetext}
\begin{eqnarray}
a(s)&=&\frac{1+v^2_\pi}{v_\pi}\left[4{\rm
Li}_2\left(\frac{1-v_\pi}{1+v_\pi}\right)+2{\rm
Li}_2\left(-\frac{1-v_\pi}{1+v_\pi}\right)-3\ln\frac{2}{1+v_\pi}
\ln\frac{1+v_\pi}{1-v_\pi}-2\ln
v_\pi\ln\frac{1+v_\pi}{1-v_\pi}\right]-\nonumber\\&&
3\ln\frac{4}{1-v^2_\pi}-4\ln v_\pi
+\frac{1}{v^3_\pi}\left[\frac{5}{4}(1+v^2_\pi)^2-2\right]\ln\frac{1+v_\pi}{1-v_\pi}
+\frac{3(1+v^2_\pi)}{2v^2_\pi}.
\end{eqnarray}
\end{widetext}
Here, $v_\pi\equiv v_\pi(s)$ is given by Eq.~(\ref{vps}), and
$${\rm Li}_2(x)=-\int_0^xdt\frac{\ln(1-t)}{t}.$$

First of all, no fit with the single $\rho(770)$ resonance
contribution, based on Eq.~(\ref{Fpifin}) in which both
$g_{\rho_2\pi\pi}$ and  $g_{\rho_3\pi\pi}$ are set to zero, is
capable of satisfactory description of all four data sets, even
with  the $\rho\omega$ mixing effect being taken into account.
Although the formula with the single resonance  works well in the
$\rho\omega$ resonance region, the curve at the far-right shoulder
of the $\rho(770)$ resonance peak does not follow the data points.

Taking into account the resonance $\rho_2$, but with the neglect
of the $\rho_3$ one, results in a rather poor fit, too. This is
because the normalization condition $F_\pi(0)=1$ reduces, in this
case, to the rather restrictive sum rule
$$
\frac{g_{\rho_1\pi\pi}}{g_{\rho_1}}+\frac{g_{\rho_2\pi\pi}}{g_{\rho_2}}=1,$$
which fixes completely the $\rho_2$ contribution to the
$e^+e^-\to\pi^+\pi^-$ reaction amplitude in a way that forbids the
successful fit. Specifically, the ratio
$g_{\rho_2\pi\pi}/g_{\rho_2}$ turns out to be too small due to the
fact that the universality condition
$g_{\rho_1\pi\pi}/g_{\rho_1}\approx1$ is satisfied for the
couplings of $\rho(770)$. See Eq.~(\ref{normFpi}). Hence,  the
$\rho_2$ resonance contribution turns out to be smaller than
necessary for reconciling the calculations with the data.  The
third resonance $\rho_3\equiv\rho(1700)$ is required in order both
to preserve the approximate universality condition and to allow a
freedom in the variation of the $\rho_2\equiv\rho(1450)$
couplings.

Free parameters, which should be determined from comparison with
the existing data \cite{snd,cmd,kloe,babar}, are the masses of the
resonances $\rho(770)$ and $\omega(782)$, the coupling constants
$g_{\rho_{1,2,3}\to\pi\pi}$ of the resonances  $\rho_{1,2,3}$ with
the $\pi^+\pi^-$ state, the coupling constants $g_{\rho_{1,2}}$
and $g_\omega$ parametrizing the $\rho_{1,2,3}$ and $\omega(782)$
leptonic decay widths [see Eq.~(\ref{gamee})], and the real part
of the polarization operator of the $\rho(770)-\omega(782)$ mixing
$\Pi^\prime_{\rho_1\omega}$. Note that $g_{\rho_3}$ is not free
but should be determined from the sum rule Eq.~(\ref{norm}). At
last, there is the parameter $a_{23}$ [see Eq.~(\ref{Piij})] that
defines Re$\Pi_{\rho_2\rho_3}$. Since we restrict our analysis to
the energy range below 1 GeV, the masses of the resonances
$\rho(1450)$ and $\rho(1700)$ are fixed to, respectively,
$m_{\rho_2}=1.45$ GeV and $m_{\rho_3}=1.7$ GeV.

\begin{figure}
\includegraphics[width=70mm]{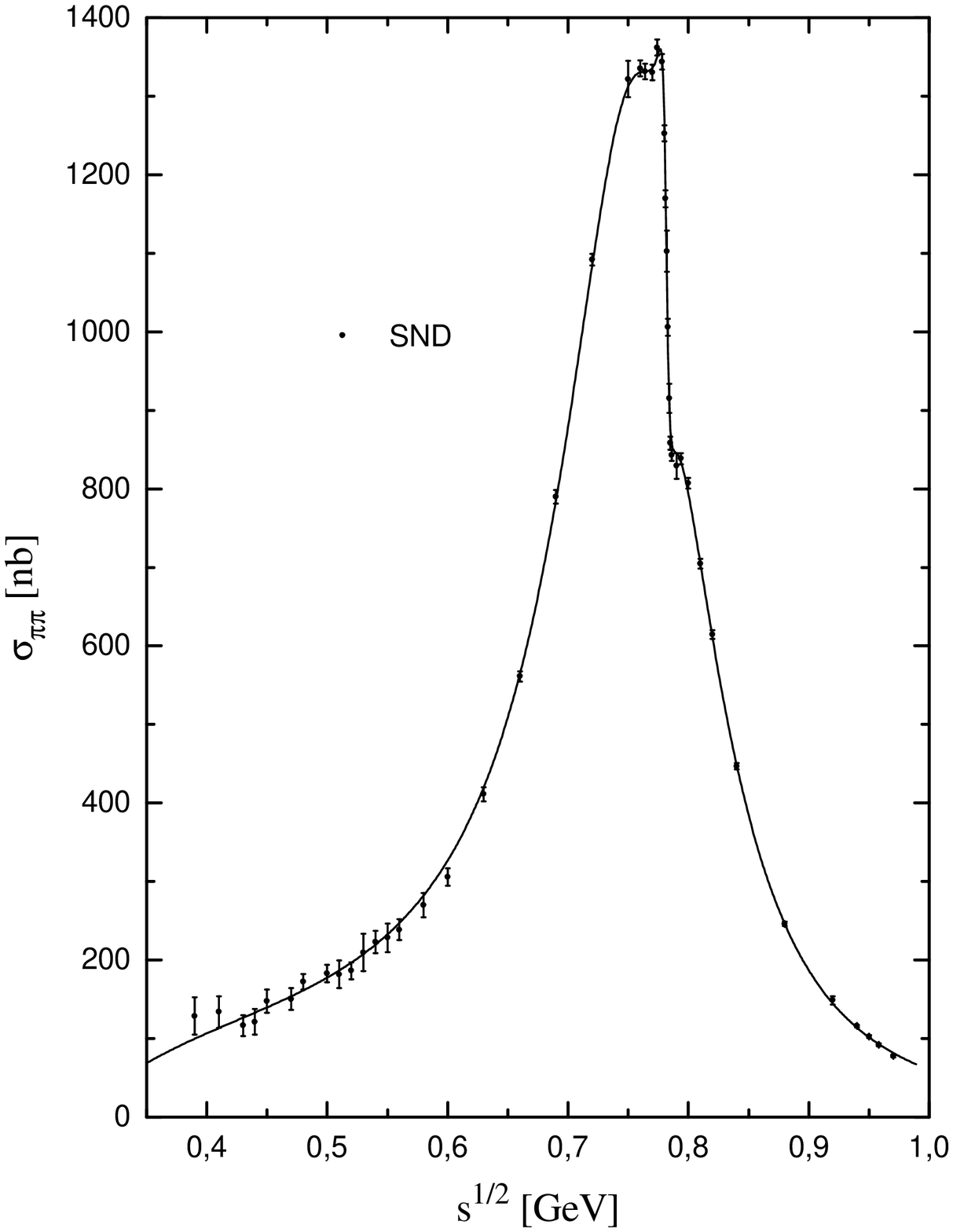}
\caption{\label{spisnd} The bare cross section,
Eq.~(\ref{sigbare}), calculated with the resonance parameters
obtained from fitting the SND data \cite{snd} listed in Table
\ref{table1}. Experimental points are from Ref.~\cite{snd}.}
\end{figure}

\begin{figure}
\includegraphics[width=70mm]{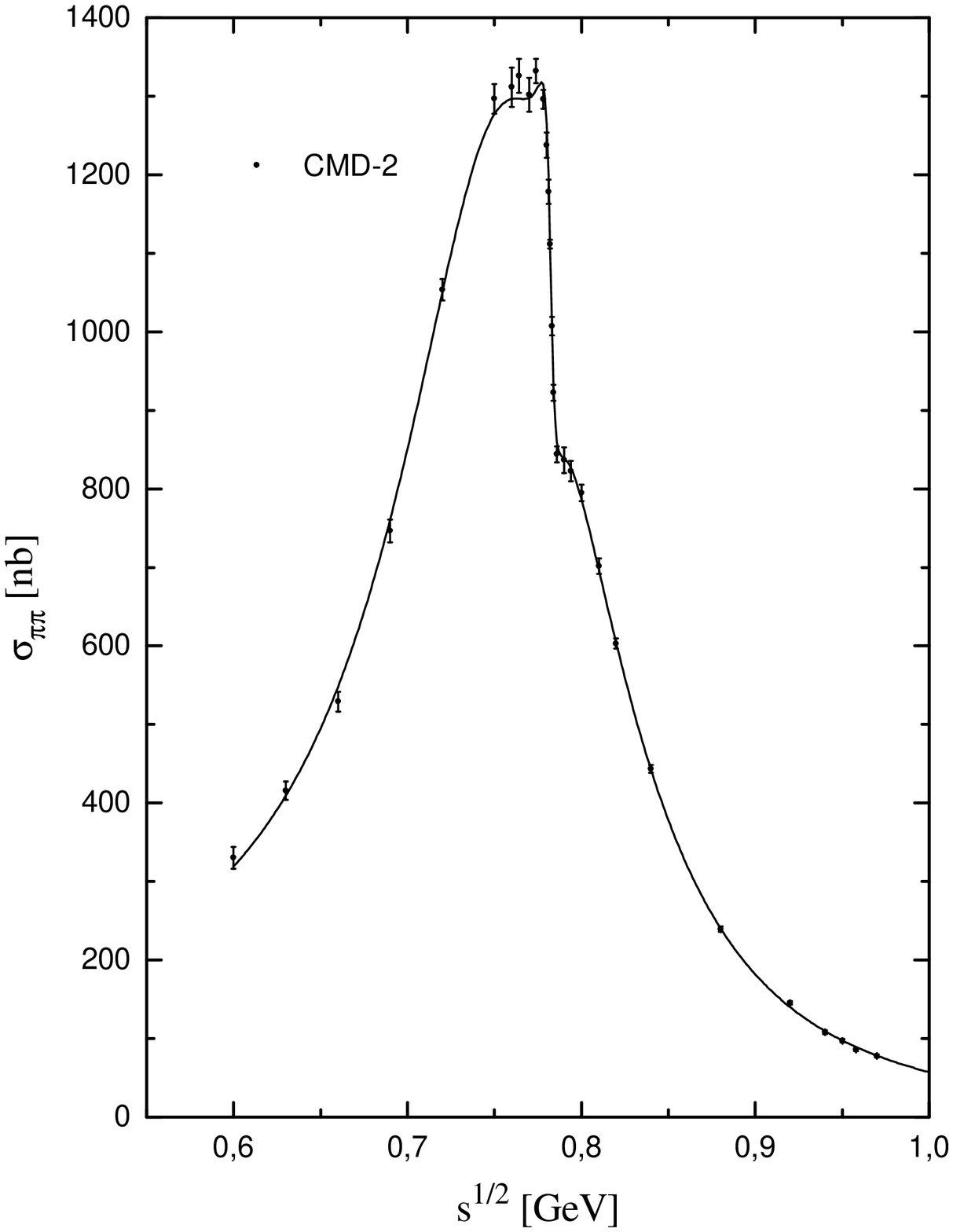}
\caption{\label{spicmd}The same as in Fig.~\ref{spisnd}, but
evaluated with the parameters obtained from fitting the CMD-2 data
\cite{cmd}. Experimental points are from Ref.~\cite{cmd}.}
\end{figure}

\begin{figure}
\includegraphics[width=70mm]{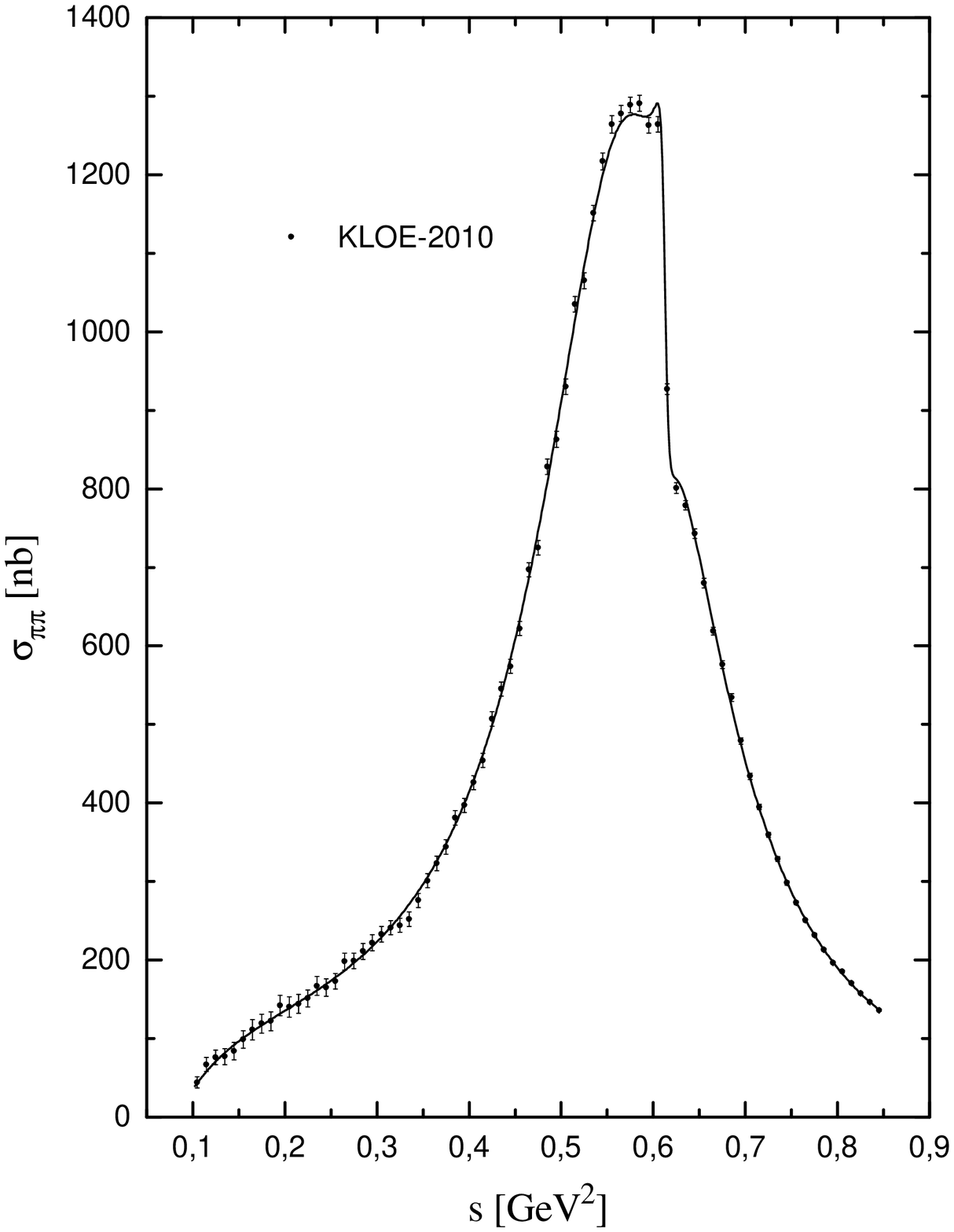}
\caption{\label{spikloe}The same as in Fig.~\ref{spisnd}, but
evaluated with the parameters obtained from fitting the KLOE-2010
data \cite{kloe}. Experimental points are from Ref.~\cite{kloe}.}
\end{figure}

\begin{figure}
\includegraphics[width=70mm]{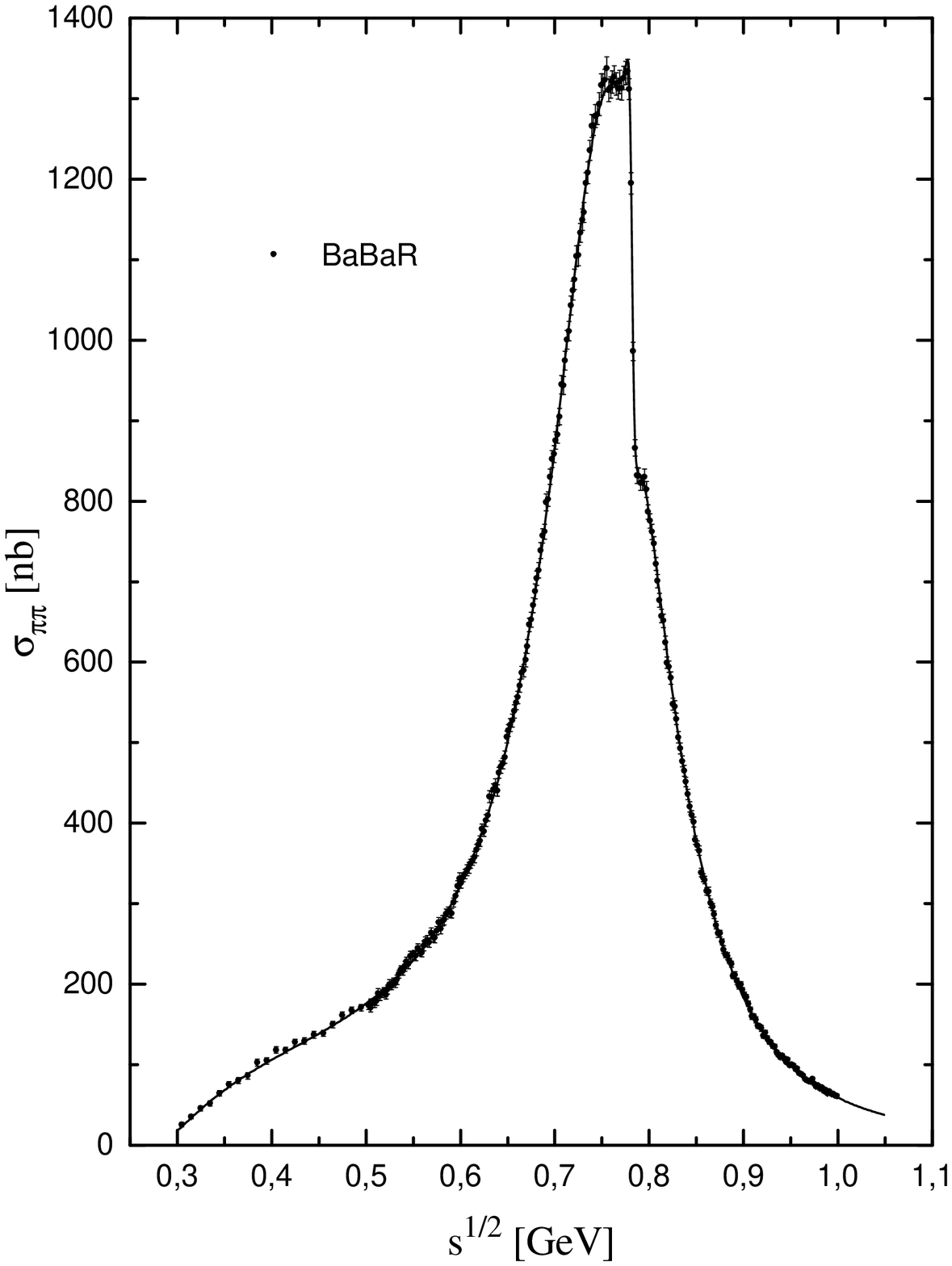}
\caption{\label{spibabar}The same as in Fig.~\ref{spisnd}, but
evaluated with the parameters obtained from fitting the BaBaR data
\cite{babar} restricted to the energies $\sqrt{s}\leq1$ GeV.
Experimental points are from Ref.~\cite{babar}.}
\end{figure}

So, the total set of free parameters is
\begin{eqnarray}
m_{\rho_1},&&g_{\rho_1\pi\pi}\mbox{, }g_{\rho_1}\mbox{,
}m_\omega\mbox{, }g_\omega\mbox{,
}\Pi^\prime_{\rho_1\omega}\mbox{, }g_{\rho_2\pi\pi}\mbox{,
}g_{\rho_2}\mbox{, }\nonumber\\&& g_{\rho_3\pi\pi}\mbox{,
}a_{23}.\label{freeparam}\end{eqnarray} Their obtained values,
found from fitting the bare cross section Eq.~(\ref{sigbare}),
side-by-side with the corresponding $\chi^2$ per number of degrees
of freedom, are listed in  Table \ref{table1} separately for the
four independent measurements of SND \cite{snd}, CMD-2 \cite{cmd},
KLOE \cite{kloe}, and the BaBaR data \cite{babar} restricted to
the low-energy range $\sqrt{s}\leq1$ GeV by the  reason explained
earlier.
\begin{table*}
\caption{\label{table1}The resonance parameters found from fitting
the data from SND \cite{snd}, CMD-2 \cite{cmd}, KLOE10
\cite{kloe}, and the BaBaR data \cite{babar} restricted to the
energies $\sqrt{s}\leq1$ GeV.}
\begin{ruledtabular}
\begin{tabular}{lllll}
 parameter&SND&CMD-2&KLOE10&BaBaR\\ \hline
 $m_{\rho_1}$
 [MeV]&$773.76\pm0.21$&$774.70\pm0.26$&$774.36\pm0.12$&$773.92\pm0.10$\\
 $g_{\rho_1\pi\pi}$&$5.798\pm0.006$&$5.785\pm0.008$&$5.778\pm0.006$&$5.785\pm0.004$\\
 $g_{\rho_1}$&$5.130\pm0.004$&$5.193\pm0.006$&$5.242\pm0.003$&$5.167\pm0.002$\\
 $m_\omega$
 [MeV]&$781.76\pm0.08$&$782.33\pm0.06$&$782.94\pm0.11$&$782.04\pm0.10$\\
 $g_\omega$&$17.13\pm0.30$&$18.43\pm0.47$&$18.27\pm0.45$&$17.05\pm0.29$\\
 $10^3\Pi^\prime_{\rho_1\omega}$
 [GeV$^2$]&$4.00\pm0.07$&$3.97\pm0.10$&$3.98\pm0.09$&$4.00\pm0.06$\\
 $g_{\rho_2\pi\pi}$&$0.71\pm0.35$&$0.79\pm0.26$&$0.019\pm0.004$&$0.21\pm0.04$\\
 $g_{\rho_2}$&$8.0\pm4.4$&$7.6\pm3.4$&$0.22\pm0.07$&$4.0\pm1.0$\\
 $g_{\rho_3\pi\pi}$&$0.20^{+1.20}_{-0.17}$&$0.76\pm0.75$&$0.055^{+0.088}_{-0.043}$&$0.011^{+0.479}_{-0.007}$\\
 $a_{23}$&$0.002\pm0.011$&$-0.016\pm0.057$&$-0.014\pm0.040$&$-0.0005\pm0.0009$\\
 $\chi^2/N_{\rm d.o.f.}$&54/35&34/19&87/65&216/260\\
\end{tabular}
\end{ruledtabular}
\end{table*}
The bare cross section evaluated with the parameters of Table
\ref{table1} is compared with the SND \cite{snd}, CMD-2
\cite{cmd}, KLOE \cite{kloe}, and BaBaR \cite{babar} data  shown
in Figs.~\ref{spisnd}, \ref{spicmd}, \ref{spikloe}, and
\ref{spibabar}, respectively.

As far as the specific values of the obtained parameters in the
Table \ref{table1} are concerned, those corresponding to  the
$\rho(770)-\omega(782)$ resonance system agree satisfactorily for
all four experiments \cite{snd,cmd,kloe,babar}. The agreement of
the coupling constants of   the resonances $\rho(1450)$ and
$\rho(1700)$ is poor but, taking into account the large
uncertainties in their determination, is not crucial. This is
justifiable, because the energy range $\sqrt{s}\leq1$ GeV is not a
proper place for extraction of the coupling constants of the above
resonances.  The widths of $\rho(1450)$ and $\rho(1700)$, in their
respective energy ranges, are known to be saturated by the
complicated final states $\rho\pi\pi$, $\omega\pi$, etc., not the
$\pi^+\pi^-$ one \cite{pdg}. Taking into account these decay modes
is necessary at energies  $\sqrt{s}>1$ GeV. Unfortunately, taking
into account the real parts of the polarization operators arising
due to the mentioned complicated states is hardly possible in
closed form. In addition, the corresponding dispersion integrals
diverge much more strongly than in the case of the $\pi^+\pi^-$
and $K\bar K$ intermediate states considered in the present work.
In the meantime, the small values of $g_{\rho_{2,3}\pi\pi}$, in
comparison with $g_{\rho_1\pi\pi}$, obtained in the present work
upon neglecting the $\rho\pi\pi$, $\omega\pi$, etc., decay modes
at $\sqrt{s}\leq1$ GeV, agree with the earlier conclusions
\cite{ach97,ach97a} inferred from the analysis in which the above
decay modes were included. Note also that $a_{23}$  is compatible
with zero.

\section{Discussion.}
\label{discussion} ~

\begin{figure}
\includegraphics[width=65mm]{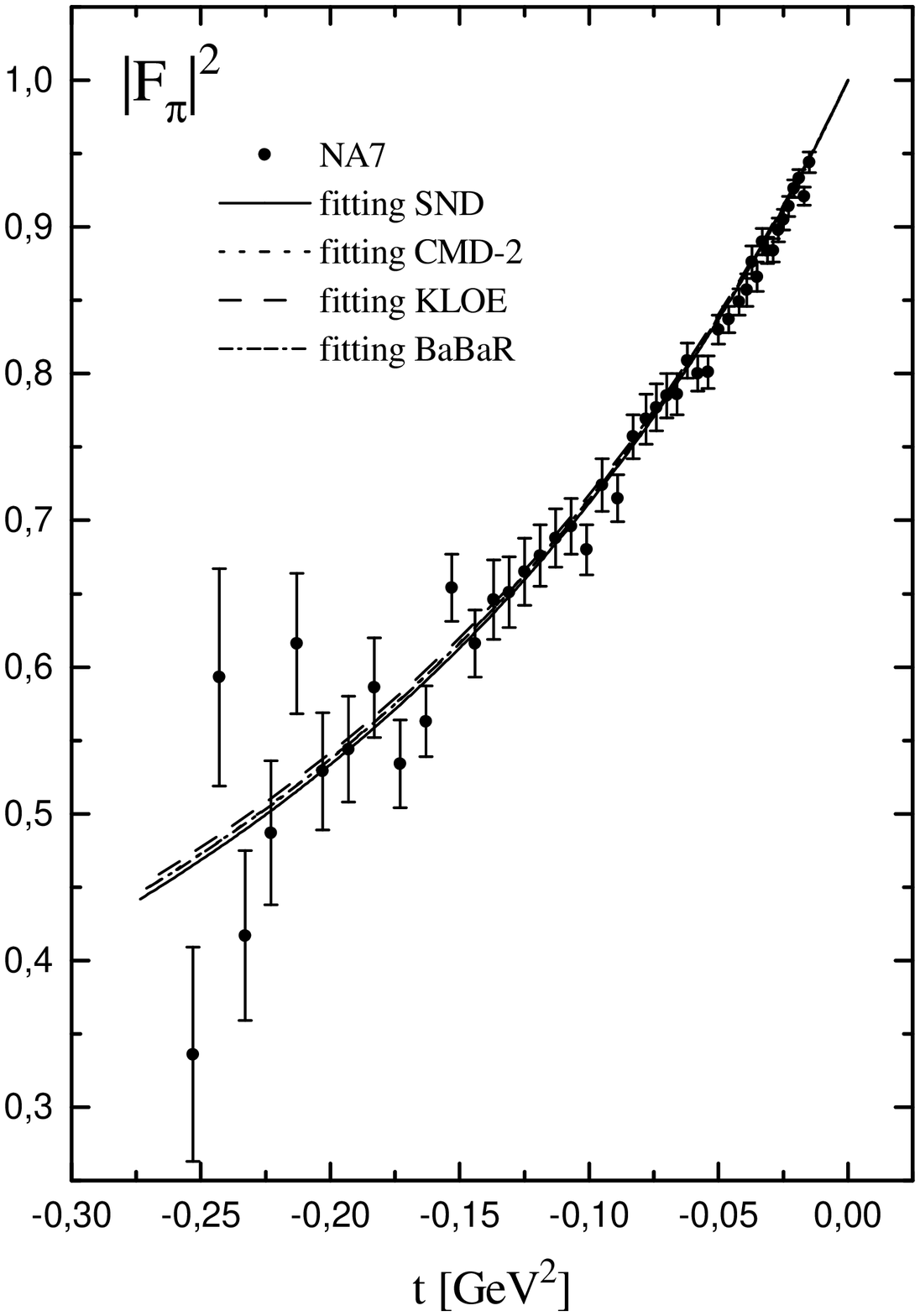}
\caption{\label{euclid}The pion form factor squared in the
spacelike region evaluated using the resonance parameters of Table
\ref{table1}. The labels of the theoretical curves correspond to
the columns of  Table \ref{table1}. The experimental data NA7 are
from Ref.~\cite{amendolia}.}
\end{figure}

An important check of the expression for the pion form factor
Eq.~(\ref{Fpifin}) and the consistency of the fits is the
continuation to the spacelike region $t<0$ accessible in the
scattering processes. To this end, one should   take the  branch
with $s<0$ in $\Pi_{\pi,K}(s,m^2_V)$ [see Eqs.~(\ref{Pipi}),
(\ref{PiKK23}), and (\ref{PiKK1})] and replace $s\to t$. Having in
mind that  the $\rho(770)-\omega(782)$ mixing in the region $t<0$
is negligibly small one can calculate $F_\pi(t)$ in this region.
The results are shown in Fig.~\ref{euclid}, where the comparison
with the NA7 data \cite{amendolia}  is presented for all four fits
considered in the present work.
We emphasize that  the data \cite{amendolia} are not included to
the fits. Hence, a good  agreement, demonstrated in
Fig.~\ref{euclid}, makes the evidence in favor of the  validity of
Eq.~(\ref{Fpifin}) for the pion form factor.

Using the resonance parameters of  Table \ref{table1}, one can
calculate, in particular, such important characteristics as the
charged pion radius $r_\pi$, defined as the square root of the
root-mean squared radius,
$$r_\pi=\sqrt{\langle r^2\rangle},$$
of the spherical symmetric electric charge distribution
\begin{eqnarray}
F_\pi({\bm q})&=&\int d^3r\rho(r)e^{i{\bm q}\cdot{\bm r}}\approx
F_\pi(0)-\nonumber\\&&\frac{{\bm
q}^2}{6}\int\rho(r)r^2d^3r=F_\pi(0)+\frac{t}{6}\langle
r^2\rangle,\label{chdistr}
\end{eqnarray}
where $t=-{\bm q}^2$. One gets
\begin{equation}
r_\pi=\sqrt{6\frac{dF_\pi(t)}{dt}}|_{t\to0}.\label{rpidef}
\end{equation}
Evaluating $r_\pi$ with the parameters of  Table \ref{table1}, one
obtains the results presented in the first row  of  Table
\ref{table2}.
\begin{table*}
\caption{\label{table2}The pion charge radius, $r_\pi$,
Eq.~(\ref{rpidef}), the renormalization constant, $Z_\rho$,
Eq.~(\ref{Zrho}),  the "physical" partial widths (with the
superscript phys), the bare ones (without the superscript), of the
decay $\rho(770)$ and $\omega(782)$, evaluated with the resonance
parameters of Table \ref{table1}.}
\begin{ruledtabular}
\begin{tabular}{lllll}
 parameter&SND&CMD-2&KLOE10&BaBaR\\ \hline
 $r_\pi$[fm]&$0.635\pm0.054$&$0.646\pm0.059$&$0.668\pm0.039$&$0.668\pm0.053$\\
 $Z_\rho$&$0.9273\pm0.0003$&$0.9277\pm0.0002$&$0.9279\pm0.0002$&$0.9277\pm0.0001$\\
 $\Gamma_{\rho_1\pi\pi}(m^2_{\rho_1})$
 [MeV]&$139.93\pm0.29$&$139.54\pm0.39$&$139.12\pm0.29$&$139.34\pm0.19$\\
 $\Gamma^{({\rm phys})}_{\rho_1\pi\pi}(m^2_{\rho_1})$
[MeV]&$150.90\pm0.31$&$150.42\pm0.42$&$149.92\pm0.31$&$150.20\pm0.20$\\
$\Gamma_{\rho_1ee}(m^2_{\rho_1})$
 [keV]&$6.56\pm0.01$&$6.41\pm0.01$&$6.29\pm0.01$&$6.47\pm0.01$\\
 $\Gamma^{({\rm phys})}_{\rho_1ee}(m^2_{\rho_1})$
 [keV]&$7.07\pm0.01$&$6.91\pm0.01$&$6.78\pm0.01$&$6.97\pm0.01$\\
$\Gamma_{\omega ee}(m^2_{\omega})$
 [keV]&$0.59\pm0.02$&$0.51\pm0.03$&$0.52\pm0.03$&$0.60\pm0.02$\\
\end{tabular}
\end{ruledtabular}
\end{table*}
For comparison, the averaged value of the pion charge radius cited
by the PDG \cite{pdg} is $r_\pi=0.672\pm0.008$ fm.

\begin{figure}
\includegraphics[width=65mm]{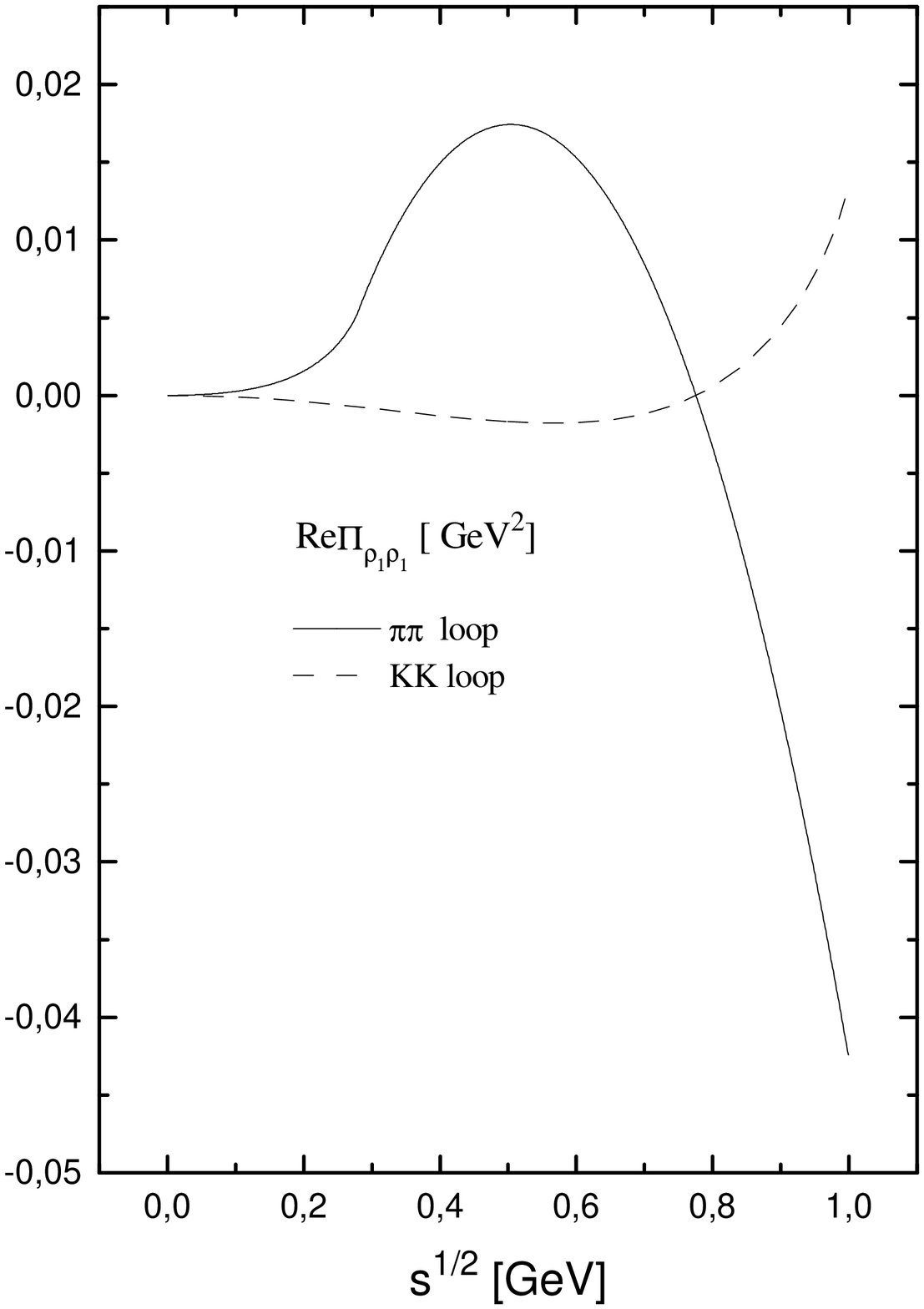}
\caption{\label{repirr}The energy dependence of ${\rm
Re}\Pi_{\rho_1\rho_1}(s)$ for both $\pi^+\pi^-$ and $K\bar K$
loops. }
\end{figure}

If one considers the single $\rho(770)$ resonance, then its
inverse propagator near $s=m^2_{\rho_1}$ can be represented as
\begin{eqnarray}
D_{\rho_1}&=&m^2_{\rho_1}-s+(m^2_{\rho_1}-s)\frac{d{\rm
Re}\Pi_{\rho_1\rho_1}(s)}{ds}\left|_{s=m^2_{\rho_1}}\right.-\nonumber\\&&
i\sqrt{s}\Gamma_{\rho_1\pi\pi}(s). \label{Dr1}
\end{eqnarray}
The behavior of ${\rm Re}\Pi_{\rho_1\rho_1}(s)$ is shown in
Fig.~\ref{repirr}.
Comparing Eq.~(\ref{Dr1}) with Eq.~(\ref{Fpisingle}) one can see
that one should make the renormalization $$g_{\rho_1\pi\pi}\to
Z_\rho^{-1/2}g_{\rho_1\pi\pi},$$ $$g_{\rho_1}\to
Z_\rho^{1/2}g_{\rho_1},$$ where
\begin{equation}
Z_\rho=1+\frac{d{\rm
Re}\Pi_{\rho_1\rho_1}(s)}{ds}\left|_{s=m^2_{\rho_1}}\right.,
\label{Zrho}\end{equation} in order to reduce Eq.~(\ref{Dr1}) to
the conveniently used form with $m_{\rho_1}$ being the physical
mass of the resonance. This results in the renormalization of the
$\pi^+\pi^-$ and $e^+e^-$ partial widths of the $\rho(770)$:
\begin{eqnarray}
\Gamma_{\rho_1\pi\pi}&\to&\Gamma^{(\rm
phys)}_{\rho_1\pi\pi}=\frac{\Gamma_{\rho_1\pi\pi}}{Z_\rho},\nonumber\\
\Gamma_{\rho_1ee}&\to&\Gamma^{(\rm
phys)}_{\rho_1ee}=\frac{\Gamma_{\rho_1ee}}{Z_\rho}.
\end{eqnarray}
The numerical values of the renormalization constant $Z_\rho$ are
given in Table \ref{table2}, side-by-side with the $\pi^+\pi^-$
and $e^+e^-$ partial widths of the $\rho(770)$. One can see that
$Z_\rho$ brings the "bare" widths (without the superscript "phys")
closer to the values $\Gamma_{\rho\pi\pi}=149.1\pm0.8$ MeV and
$\Gamma_{\rho e e}=7.04\pm0.06$ keV cited in the Review of
Particle Physics \cite{pdg}.

\begin{figure}
\includegraphics[width=65mm]{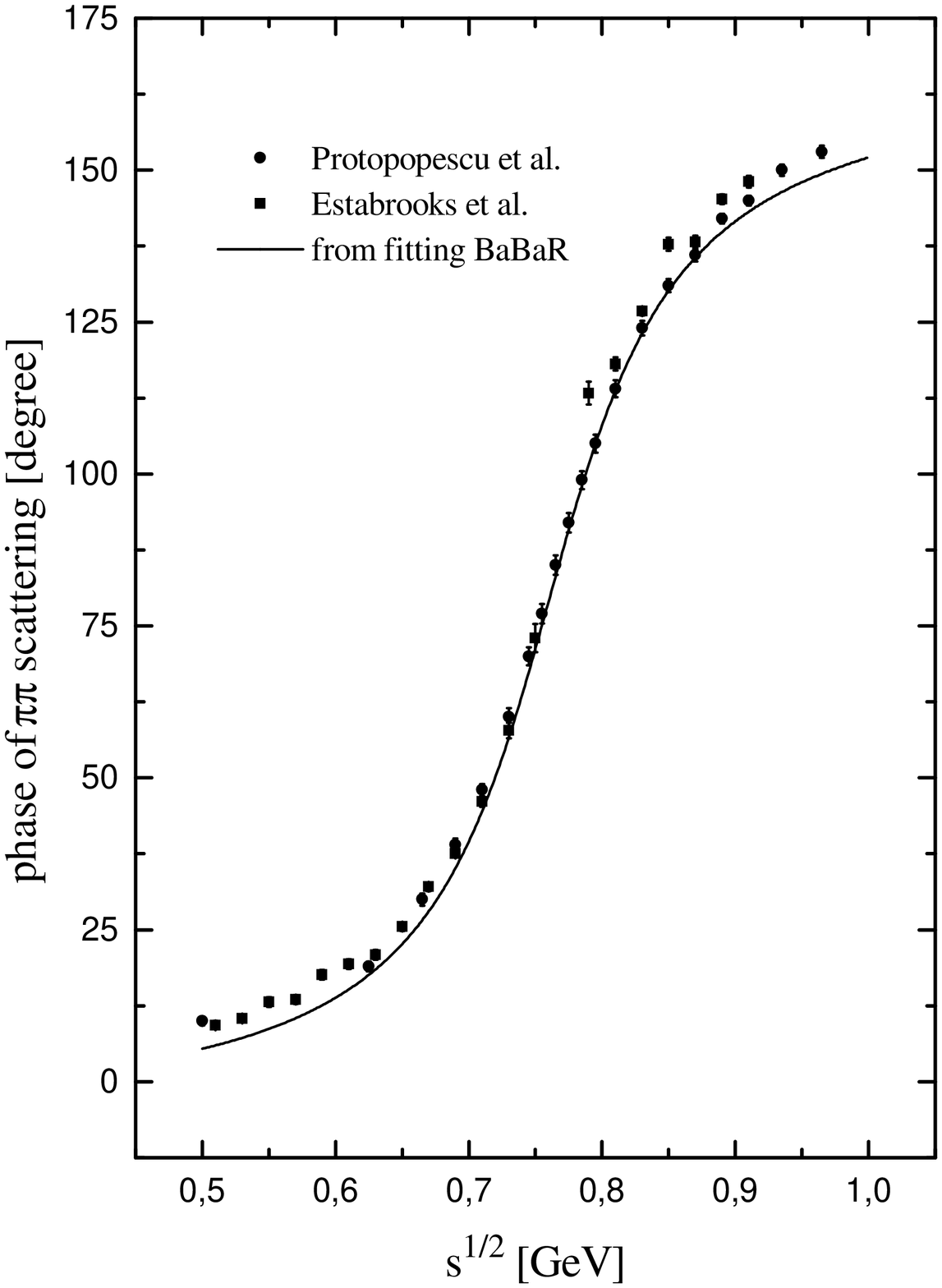}
\caption{\label{phase}The phase shift $\delta^1_1$ of $\pi\pi$
scattering. The data are, respectively, Protopopescu et al.
\cite{proto} and Estabrooks et al. \cite{estab}. The curves
corresponding to the parameters obtained from fitting the SND,
CMD-2, and KLOE data are not shown because they coincide with the
curve evaluated using the parameters from the fit of the BaBaR
data, shown here.}
\end{figure}

Another important characteristic of the low-energy hadronic
physics is the phase shift $\delta^1_1$ of $\pi\pi$ scattering in
the vector-isovector channel with the quantum numbers of
$\rho(770)$. At energies below the $\omega\pi$ and $K\bar K$
production thresholds, $\delta^1_1$ is given by the phase of the
pion form factor
\begin{equation}
\delta^1_1=\arctan\frac{{\rm Im}F_\pi}{{\rm
Re}F_\pi},\label{del11}\end{equation}where $F_\pi$ is given by
Eq.~(\ref{Fpifin}) upon neglecting the contribution of
$\rho\omega$ mixing $\propto\Pi_{\rho_1\omega}$. The plot of
$\delta^1_1$, obtained using parameters extracted from fitting the
low-energy portion of the BaBaR data \cite{babar}, is shown in
Fig.~\ref{phase}, where the comparison with the data
\cite{proto,estab} is presented. Note that the resonance
parameters, extracted from three other sets of data
\cite{snd,cmd,kloe}, result in the curves for $\delta^1_1$
coincident with that shown in Fig.~\ref{phase}. Having in mind
that the data on the phase shift were not included in the fits,
the agreement of the calculated $\delta^1_1$ with the measured one
is satisfactory.

\section{Conclusion}
\label{conclusion}
~

It is shown that the new formula for $F_\pi(s)$,
Eq.~(\ref{Fpifin}), gives a good description of the latest
experimental data \cite{snd,cmd,kloe, babar} on the production of
the $\pi^+\pi^-$ pair in $e^+e^-$ annihilation at $\sqrt{s}<1$
GeV. In this low-energy domain, one can restrict oneself by the
contribution of the $\pi^+\pi^-$ and $K\bar K$ loops to both
diagonal and nondiagonal polarization operators. In principle,
other intermediate states could be  taken into account, at least
numerically. However, heavier isovector resonances $\rho(1450)$
and $\rho(1700)$ are known to have other decay modes besides
$\pi^+\pi^-$ and $K\bar K$, such as $\omega\pi$, $a_1\pi$ etc. The
treatment should include the energies $\sqrt{s}\leq2$ GeV where
the coupling constants with the above states could be determined.
No data exist on these decay modes of the quality comparable with
the $\pi^+\pi^-$ data \cite{snd,cmd,kloe, babar}. Hence, at
present, the restriction to the domain $\sqrt{s}<1$ GeV and to the
pseudoscalar loops seems justifiable.

\begin{acknowledgements}
We are grateful to M.~N.~Achasov for numerous discussions which
stimulated the present work.
\end{acknowledgements}

\appendix
\section{The finite width and the resonance mixing} ~

Some details necessary for  taking into account the finite width
effects and the resonance mixing are given in this Appendix.  The
meaning of the diagonal polarization operator $\Pi_{RR}(s)$ is
that it modifies the inverse bare propagator of the resonance $R$
with the mass $m_R$, $D^{(0)}_R(s)\equiv D^{(0)}_R=m^2_R-s$, in
the following way:
\begin{eqnarray*}
\frac{1}{D_R(s)}&=&\frac{1}{D^{(0)}_R}+\frac{1}{D^{(0)}_R}\Pi_{RR}(s)\frac{1}
{D^{(0)}_R}+\nonumber\\&&\frac{1}{D^{(0)}_R}\Pi_{RR}(s)\frac{1}{D^{(0)}_R}\Pi_{RR}(s)
\frac{1}{D^{(0)}_R}+\cdots=\nonumber\\&&\frac{1}{D^{(0)}_R-\Pi_{RR}(s)}.\end{eqnarray*}
In particular, this  formula takes into account the finite width
effects
\begin{eqnarray}
D_R(s)&=&m^2_R-s-{\rm
Re}\Pi_{RR}(s)-i\sqrt{s}\Gamma_{R\pi\pi}(s).\label{DR}\end{eqnarray}

In principle, the mixing of the isovector resonances $\rho(770)$,
$\rho(1450)$, and $\rho(1700)$ can be strong, especially because
of the common decay modes, for example, the $\pi^+\pi^-$ one. It
can be taken into account in the field-theory-inspired approach
based on summing to all orders of the loop corrections to the bare
propagators of vector mesons \cite{ach84,ach97,ach97a,ach02}. The
term "bare" means that the propagators are not distorted by the
mixing. The scheme can be demonstrated by taking the two-resonance
mixing as an example \cite{ach02}. It   reduces in this case to
the following replacements:
\begin{eqnarray*}
\frac{1}{D_R}&\to&\frac{1}{D_R}+\frac{1}{D_R}\Pi_{RR^\prime}\frac{1}{D_{R^\prime}}
\Pi_{RR^\prime}\frac{1}{D_R}+\cdots=\nonumber\\&&
\frac{D_{R^\prime}}{D_RD_{R^\prime}-\Pi^2_{RR^\prime}}\equiv\left(G^{-1}\right)_{RR},\nonumber\\
\frac{1}{D_{R^\prime}}&\to&\frac{1}{D_{R^\prime}}+\frac{1}{D_{R^\prime}}\Pi_{RR^\prime}\frac{1}{D_R}
\Pi_{RR^\prime}\frac{1}{D_{R^\prime}}+\cdots=\nonumber\\&&
\frac{D_R}{D_RD_{R^\prime}-\Pi^2_{RR^\prime}}\equiv\left(G^{-1}\right)_{R^\prime R^\prime},\nonumber\\
\frac{\Pi_{RR^\prime}}{D_RD_{R^\prime}}&\to&\frac{\Pi_{RR^\prime}}{D_RD_{R^\prime}}+
\frac{(\Pi_{RR^\prime})^3}{(D_RD_{R^\prime})^2}+\cdots=\nonumber\\&&
\frac{\Pi_{RR^\prime}}{D_RD_{R^\prime}-\Pi^2_{RR^\prime}}
\equiv(G^{-1})_{RR^\prime}.
\end{eqnarray*}
The matrix
$$G=\left(%
\begin{array}{cc}
  D_R & -\Pi_{RR^\prime} \\
  -\Pi_{RR^\prime} & D_{R^\prime} \\
\end{array}%
\right)
$$is the matrix of inverse propagators in the two-resonance case.
Let us take for a moment just this  case, $R=\rho_1$ and
$R^\prime=\rho_2$, in order to clarify the effect of the mixing on
the resonance position. Neglecting for a moment  the $\rho\omega$
mixing which is taken into account below, one can write the pion
form factor as
\begin{eqnarray}
F_\pi&=&(g_{\gamma\rho_1},g_{\gamma\rho_2})\left(%
\begin{array}{cc}
  D_{\rho_2} &\Pi_{\rho_1\rho_2}  \\
  \Pi_{\rho_1\rho_2} & D_{\rho_1} \\
\end{array}%
\right)\left(%
\begin{array}{c}
  g_{\rho_1\pi\pi} \\
  g_{\rho_2\pi\pi} \\
\end{array}%
\right)\times\nonumber\\&&
\frac{1}{D_{\rho_1}D_{\rho_2}-\Pi^2_{\rho_1\rho_2}}.
\label{Fpi2r}\end{eqnarray} In the vicinity of the $\rho_1$
resonance position, $s\to m^2_{\rho_1}$, Eq.~(\ref{Fpi2r}) can be
represented in the form
\begin{widetext}
\begin{eqnarray}
F_\pi(s)\approx\frac{g_{\gamma\rho_1}g_{\rho_1\pi\pi}}{m^2_{\rho_1}-s-
\Pi_{\rho_1\rho_1}(s)-\frac{\Pi^2_{\rho_1\rho_2}(m^2_{\rho_1})}{m^2_{\rho_2}-m^2_{\rho_1}
-\Pi_{\rho_2\rho_2}(m^2_{\rho_1})}}, \label{Fpiappr}\end{eqnarray}
\end{widetext}
where, in accord with the adopted  definition,
Re$\Pi_{\rho_1\rho_1}(m^2_{\rho_1})=0$. One can see from
Eq.~(\ref{Fpiappr}) that  there is a  shift in the $\rho_1$
resonance peak position due to the mixing of $\rho_1$ with the
resonance $\rho_2$:
\begin{eqnarray}
\Delta m^2_{\rho_1}&=&-{\rm
Re}\frac{\Pi^2_{\rho_1\rho_2}(m^2_{\rho_1})}{m^2_{\rho_2}-m^2_{\rho_1}
-\Pi_{\rho_2\rho_2}(m^2_{\rho_1})}\approx\nonumber\\&&-\frac{{\rm
Re}\left[\Pi^2_{\rho_1\rho_2}(m^2_{\rho_1})\right]}{m^2_{\rho_2}-m^2_{\rho_1}},
\label{Deltamrho}
\end{eqnarray}
where we neglect $\Pi_{\rho_2\rho_2}(m^2_{\rho_1})$ in comparison
with the mass difference squared $m^2_{\rho_2}-m^2_{\rho_1}$.
Indeed, using the plots in Fig.~\ref{repirr}, the relation
$$\Pi_{\rho_2\rho_2}=\left(\frac{g_{\rho_2\pi\pi}}{g_{\rho_1\pi\pi}}\right)^2
\Pi_{\rho_1\rho_1},$$ Eq.~(\ref{ImPiii}), Eq.~(\ref{Piii}),  and
$g_{\rho_2\pi\pi}\approx0.8$ (see Table \ref{table1}), one obtains
the estimate
$$\frac{\Pi_{\rho_2\rho_2}(m^2_{\rho_1})}{m^2_{\rho_2}-
m^2_{\rho_1}}\lesssim(0.2+1.5i) \times10^{-3}.$$ In the case of
the well-studied resonance $\rho_1=\rho(770)$, it is natural to
expect that the visible peak position with a good accuracy
coincides with the bare mass $m_{\rho_1}$. This follows from the
definition Re$\Pi_{\rho_1\rho_1}(m^2_{\rho_1})=0$ adopted in the
present work. In order to preserve the above coincidence, the
natural demand is to set Re$\Pi_{\rho_1\rho_2}=0$. Since, in
Eq.~(\ref{Deltamrho}), Re$\Pi^2_{\rho_1\rho_2}=({\rm
Re}\Pi_{\rho_1\rho_2})^2-({\rm Im}\Pi_{\rho_1\rho_2})^2$, then, to
be precise, some mass shift survives which is equal to
$$\Delta m_{\rho_1}\approx\frac{m_{\rho_1}\Gamma^2_{\rho_1\pi\pi}(m^2_{\rho_1})}
{2(m^2_{\rho_2}-m^2_{\rho_1})}\left(\frac{g_{\rho_2\pi\pi}}{g_{\rho_1\pi\pi}}\right)^2.$$
However, even in the worse case $g_{\rho_2\pi\pi}=0.8$ (see Table
\ref{table1}, where the magnitudes of the coupling constants
extracted from the specific fits are given), this shift is
estimated at the level of 0.1 MeV. This estimate  falls within the
errors of $m_{\rho_1}$, quoted in  Table \ref{table1}. Having in
mind the three-resonance case, we set Re$\Pi_{\rho_1\rho_3}=0$.
Such a type of justification is not applicable for the poorly
studied resonances $\rho_2=\rho(1450)$ and $\rho_3=\rho(1700)$;
hence, the parameter $a_{23}$ fixing Re$\Pi_{\rho_2\rho_3}$
remains free.

The generalization to the case of three (and any number of)
resonances $\rho_1,\rho_2$, and $\rho_3$ is straightforward. The
matrix of inverse propagators is given by Eq.~(\ref{G}). The
matrix of propagators is
$$
G^{-1}=\frac{1}{\Delta}\left(%
\begin{array}{ccc}
  g_{11} & g_{12} & g_{13}  \\
  g_{12} & g_{22} & g_{23}  \\
  g_{13} & g_{23} & g_{33}  \\
  \end{array}%
\right),
$$
where
\begin{eqnarray}
g_{11}&=&D_{\rho_2}D_{\rho_3}-\Pi^2_{\rho_2\rho_3},\nonumber\\
g_{22}&=&D_{\rho_1}D_{\rho_3}-\Pi^2_{\rho_1\rho_3},\nonumber\\
g_{33}&=&D_{\rho_1}D_{\rho_2}-\Pi^2_{\rho_1\rho_3},\nonumber\\
g_{12}&=&D_{\rho_3}\Pi_{\rho_1\rho_2}+\Pi_{\rho_1\rho_3}\Pi_{\rho_2\rho_3},\nonumber\\
g_{13}&=&D_{\rho_2}\Pi_{\rho_1\rho_3}+\Pi_{\rho_1\rho_2}\Pi_{\rho_2\rho_3},\nonumber\\
g_{23}&=&D_{\rho_1}\Pi_{\rho_2\rho_3}+\Pi_{\rho_1\rho_2}\Pi_{\rho_1\rho_3},\nonumber\\
\Delta&\equiv&{\rm
det}G=D_{\rho_1}D_{\rho_2}D_{\rho_3}-2\Pi_{\rho_1\rho_2}\Pi_{\rho_1\rho_3}\Pi_{\rho_2\rho_3}
-\nonumber\\&&D_{\rho_1}\Pi^2_{\rho_2\rho_3}-
D_{\rho_2}\Pi^2_{\rho_1\rho_3}-D_{\rho_3}\Pi^2_{\rho_1\rho_2}.
\label{gij}\end{eqnarray} Note that, deep in the spacelike domain,
the quantity $1/\Delta$ and, as a consequence, the pion form
factor have a pole at $\sqrt{-t}=$87, 82, 97, and 95 GeV, when
evaluated with the resonance parameters obtained from the fit of,
respectively, SND \cite{snd}, CMD-2 \cite{cmd}, KLOE \cite{kloe},
and BaBaR \cite{babar} data. This pole is the analog of the famous
Landau pole.

In addition to the strong mixing between the isovector resonances,
one should include also the isovector-isoscalar
$\rho_i-\omega(782)$ mixing arising due to small  G-parity
breaking. Then, the matrix of inverse propagators can be written
in the form
\begin{equation}
G_{\rm tot}=\left(%
\begin{array}{cccc}
  D_{\rho_1} & -\Pi_{\rho_1\rho_2} & -\Pi_{\rho_1\rho_3} & -\Pi_{\rho_1\omega} \\
  -\Pi_{\rho_1\rho_2} & D_{\rho_2} & -\Pi_{\rho_2\rho_3} & -\Pi_{\rho_2\omega} \\
  -\Pi_{\rho_1\rho_3} & -\Pi_{\rho_2\rho_3} & D_{\rho_3} & -\Pi_{\rho_3\omega} \\
  -\Pi_{\rho_1\omega} & -\Pi_{\rho_2\omega} & -\Pi_{\rho_3\omega} & D_\omega \\
\end{array}%
\right)
\end{equation}
In this case, the pion form factor is written as follows:
\begin{equation}
F_\pi(s)=(g_{\gamma\rho_1},g_{\gamma\rho_2},g_{\gamma\rho_3},g_{\gamma\omega})G^{-1}\left(\begin{array}{c}
                                                                            g_{\rho_1\pi\pi} \\
                                                                            g_{\rho_2\pi\pi} \\
                                                                            g_{\rho_3\pi\pi} \\
                                                                            g_{\omega\pi\pi} \\
                                                                          \end{array}\right),
                                                                          \label{Fpi123}
                                                                          \end{equation}
The coupling constant $g_{\omega\pi\pi}$ describes the direct
$\omega\to\pi^+\pi^-$ transition arising due to the violation of
G-parity conservation side-by-side with the mixing mechanism.
However, it is known \cite{gourdin} that,  since the $\pi^+\pi^-$
channel dominates the $\rho_1$ decay width, $g_{\omega\pi\pi}$ is
almost canceled in the effective $\omega\to\pi^+\pi^-$ transition
amplitude due to the compensation among  imaginary parts of
$\Pi_{\rho_1\omega}$ and the inverse $\rho_1$ propagator. Indeed,
allowing for  both the mixing and direct transition, one can write
the effective $\omega\pi\pi$ coupling constant in the form
\begin{widetext}
\begin{eqnarray}
g_{\omega\pi\pi}^{({\rm
eff})}&\approx&g_{\omega\pi\pi}-\frac{({\rm
Re}\Pi_{\rho_1\omega}+i{\rm
Im}\Pi_{\rho_1\omega})g_{\rho_1\pi\pi}}{m^2_\omega-m^2_{\rho_1}-i\sqrt{s}(\Gamma_\omega-\Gamma_{\rho_1\pi\pi})}
=\frac{1}{m^2_\omega-m^2_{\rho_1}-i\sqrt{s}(\Gamma_\omega-\Gamma_{\rho_1\pi\pi})}
\left\{g_{\omega\pi\pi}\left[m^2_\omega-m^2_{\rho_1}-\right.\right.\nonumber\\&&\left.\left.
i\sqrt{s}(\Gamma_\omega-\Gamma_{\rho_1\pi\pi})\right]-
g_{\rho_1\pi\pi}\left[{\rm Re}\Pi_{\rho_1\omega}+i\left({\rm
Im}\widetilde{\Pi}_{\rho_1\omega}+\sqrt{s}\frac{g_{\omega\pi\pi}}
{g_{\rho_1\pi\pi}}\Gamma_{\rho_1\pi\pi}\right)\right]\right\}\approx\nonumber\\&&
-\frac{({\rm Re}\Pi_{\rho_1\omega}+i{\rm
Im}\widetilde{\Pi}_{\rho_1\omega})g_{\rho_1\pi\pi}}
{m^2_\omega-m^2_{\rho_1}-i\sqrt{s}(\Gamma_\omega-\Gamma_{\rho_1\pi\pi})},
\label{gom2pi}
\end{eqnarray}
\end{widetext}
where ${\rm Im}\widetilde{\Pi}_{\rho_1\omega}$ differs from ${\rm
Im}\Pi_{\rho_1\omega}$ by the absence of the term $\propto
g_{\omega\pi\pi}$. Hence, one can safely neglect the coupling
constant  $g_{\omega\pi\pi}$. This circumstance  was not properly
accounted for in our earlier work, Ref.~\cite{ach97}. The
isovector-isoscalar type of weak mixing is essential only for the
$\rho(770)-\omega(782)$ system because it is enhanced due to the
small mass difference of these resonances. See Eq.~(\ref{gom2pi}).
As for other isovector-isoscalar mixings $\rho(1450)-\omega(782)$
and $\rho(1700)-\omega(782)$, there is no enhancement, due to the
mass proximity, and    one can  neglect $\Pi_{\rho_{2,3}\omega}$.
Taking the latter assumption into account and allowing for the
$\rho_1\omega$ mixing to first order, one can approximate the
propagator matrix $G^{-1}$ in Eq.~(\ref{Fpi123}) by the expression
$$
G^{-1}_{\rm tot}\approx\frac{1}{\Delta}\left(%
\begin{array}{cccc}
  g_{11} & g_{12} & g_{13} & \frac{g_{11}\Pi_{\rho_1\omega}}{D_\omega} \\
  g_{12} & g_{22} & g_{23} & \frac{g_{12}\Pi_{\rho_1\omega}}{D_\omega} \\
  g_{13} & g_{23} & g_{33} & \frac{g_{13}\Pi_{\rho_1\omega}}{D_\omega} \\
  \frac{g_{11}\Pi_{\rho_1\omega}}{D_\omega} & \frac{g_{12}\Pi_{\rho_1\omega}}{D_\omega} &
  \frac{g_{11}\Pi_{\rho_1\omega}}{D_\omega} & \frac{\Delta}{D_\omega} \\
\end{array}%
\right),
$$where the $g_{ij}$ and $\Delta$ are given by
Eq.~(\ref{gij}). The final approximate expression for the pion
form factor $F_\pi\equiv F_\pi(s)$  given by Eq.~(\ref{Fpifin}) is
obtained by inserting  this approximate expression to
Eq.~(\ref{Fpi123}) and by neglecting the coupling constant of the
direct decay $g_{\omega\pi\pi}$.


\end{document}